# The Geospatial Characteristics of a Social Movement Communication Network

**Michael D. Conover\*, Clayton Davis, Emilio Ferrara, Karissa McKelvey, Filippo Menczer, Alessandro Flammini**

Center for Complex Networks and Systems Research, School of Informatics and Computing, Indiana University, Bloomington, Indiana, United States of America

## Abstract

Social movements rely in large measure on networked communication technologies to organize and disseminate information relating to the movements' objectives. In this work we seek to understand how the goals and needs of a protest movement are reflected in the geographic patterns of its communication network, and how these patterns differ from those of stable political communication. To this end, we examine an online communication network reconstructed from over 600,000 tweets from a thirty-six week period covering the birth and maturation of the American anticapitalist movement, Occupy Wall Street. We find that, compared to a network of stable domestic political communication, the Occupy Wall Street network exhibits higher levels of locality and a hub and spoke structure, in which the majority of non-local attention is allocated to high-profile locations such as New York, California, and Washington D.C. Moreover, we observe that information flows across state boundaries are more likely to contain framing language and references to the media, while communication among individuals in the same state is more likely to reference protest action and specific places and times. Tying these results to social movement theory, we propose that these features reflect the movement's efforts to mobilize resources at the local level and to develop narrative frames that reinforce collective purpose at the national level.

**Funding:** The authors gratefully acknowledge support from the National Science Foundation [http://www.nsf.gov/] (grant CCF-1101743), DARPA [http://www.darpa.mil/] (grant W911NF-12-1-0037), and the McDonnell Foundation [http://www.jsmf.org/]. Any opinions, findings, and conclusions or recommendations expressed in this material are those of the author(s) and do not necessarily reflect the views of the funding agencies. The funders had no role in study design, data collection and analysis, decision to publish, or preparation of the manuscript.

**Competing Interests:** The authors have declared that no competing interests exist.

* E-mail: midconov@indiana.edu

## Introduction

One of the most prominent American political movements of the past thirty years, Occupy Wall Street ('Occupy') is remarkable in the extent to which social media played a central role in its development and organization [1,2]. In this study, we examine how the needs and constraints of social movements are reflected in the geospatial characteristics and information sharing practices of Twitter users engaged in communication about the Occupy movement. Specifically, we focus on the geographic distribution of these users and the ways in which the relationships among them diverge from those of users contributing to the two most popular streams for stable political discourse in the United States, 'Top Conservatives on Twitter' and 'Progressives 2.0.'

The organizing forces underlying successful social movements have been studied extensively by sociologists and political scientists [3–7]. From this body of work common themes have emerged, include the problems of resource mobilization and collective framing, which together constitute two of the core issues any social movement must address in order to effect social or political change. Resource mobilization refers to the process by which a social movement must marshal the financial, material, and human resources required to sustain its activities [8]. Collective framing is a process whereby the constituents of a social movement, through formal or informal processes, come to establish the narratives, language, and imagery that capture the essential features of the movement's purpose and struggle [9]. Effective framing helps to foster a sense of community and engagement, and can be a powerful response to countervailing social pressures from establishment organizations [10].

Here we study Occupy Wall Street, a social movement focused on issues relating to the uneven distribution of wealth, social inequality, corporate greed, and the regulation of major financial institutions. Since the first protest on September 17th, 2011, a major feature of the movement has been the long-term physical occupation of high-visibility encampments, often found in parks, banks, libraries and foreclosed homes. As a result, the Occupy movement requires substantial supporting infrastructure, including housing and sanitation facilities, as well as access to communication technologies. In spite of this, Occupy has sustained a lasting presence in American cities including New York City, Oakland, Washington, D.C., and Boston, which also represent key loci of decision making and protest activity [1,2]. Under the Occupy model, proposals are brought to a vote before a general assembly, a form of direct democracy in which any participant is free to comment or vote on any proposal under consideration. The most prominent among these organizational structures is the New York City General Assembly, which has been responsible for producing policy and key narrative frames such as the popular protest slogan, "We are the 99%," which references the disproportionate concentration of wealth among the top 1% of the world's population [11].

Social media have played a prominent role in facilitating communication and coordination throughout the development of the Occupy Wall Street movement. For example, the first call to action in the Canadian anticapitalist magazine 'AdBusters' used the Twitter 'hashtag' #occupywallstreet as one of just ten words featured in a full-page ad. Ever since, the Twitter platform has been used extensively by movement participants [2], with #ows being one of the hundred most popular hashtags on Twitter for the year 2011. In addition to Occupy, Twitter has also played a prominent role in several foreign social movements, most notably in the Egyptian revolutionary protests of 2011 [12–14].

In this work, we seek to understand the relationship between the geospatial dimensions of social movement communication networks and the organizational pressures facing such movements. To accomplish this, we use a state-of-the-art location inference technique to model relationships among users as a weighted directed network of communication flows between states, in which the weight of each edge corresponds to the volume of traffic between pairs of locations. Using this framework we investigate three distinct relationships: attention allocation and proximity to on-the-ground events, resource mobilization and localized information sharing, and the role of collective framing in long-distance communication.

With respect to the issue of attention allocation, we find that compared to stable domestic political communication the Occupy Wall Street movement exhibits very high levels of geographic concentration, with users in New York, California, and Washington D.C. producing more than half of all retweeted content. Aside from these hubs, however, we find that the appeal of content relating to Occupy Wall Street has a disproportionately local audience. With extended, high profile encampments and large-scale protest action playing central roles in the Occupy movement, we propose that this structural feature reflects the importance of mobilizing human resources at the local level.

Finally, we report on evidence indicating that the content of communication at the national level is distinct from the content of communication among users in the same state. Comparing intrastate versus interstate communication, we find that the terms most overrepresented in interstate communication relate to the movement's core framing language and the news media, while the terms most overrepresented in local communication reference physical places, protest action, and specific times. These results support the hypothesis that local-level communication activity is driven by the challenge of resource mobilization, while long-distance communication is more strongly associated with collective framing processes.

## Materials and Methods

### Twitter Platform

Twitter is a popular social networking and microblogging site extensively explored in recent literature [15–21]. Among others, it has been used to study influence and credibility [22–26], social structure [27–29] and to monitor users' sentiment [30–33]. Twitter users can post 140-character messages containing text and hyperlinks, called *tweets*, and interact with one another in a variety of ways. Communication on Twitter is characterized by directed, non-reciprocal social links that allow users to subscribe to the stream of content produced by another user. The content produced by every user an individual follows is aggregated into a single streaming feed, from which an individual can selectively rebroadcast content to his or her followers by choosing to retweet it. In this way, a retweet serves to broaden the potential audience of a piece of content, and signifies that information has been transmitted between two individuals. Hashtags, short tokens prepended with a pound sign (e.g. #taxes or #obama), constitute another important feature of the platform, and allow the content produced by many individuals to be aggregated into a custom, topic-specific stream including all tweets containing a given token.

### Data

The analysis described in this article relies on data collected from the Twitter 'gardenhose' streaming API between July $3^{rd}$, 2011 and March $12^{th}$, 2012 – a nine month period including the birth and maturation of the Occupy Wall Street movement. The gardenhose provides an approximately 10% sample of the entire Twitter stream in a machine-readable format. Gardenhose tweets include useful metadata, among them a unique tweet identifier, the content of the tweet (including hashtags and hyperlinks), a timestamp, the username of the account that produced the tweet, a free text 'location' string associated with the originating user's profile, and for retweets, the account names of the other users associated with the tweet. Tweets from geolocation-enabled mobile devices also report latitude/longitude coordinates, however the incidence rate of tweets with this data is not enough to be useful as a feature in general.

To isolate a representative sample of Occupy Wall Street content we flagged for collection any tweet containing hashtags associated with the Occupy movement, including #ows and #occupy{*} (e.g. #occupywallst, #occupyboston, etc.). To provide a baseline against which to compare our observations, we also extracted content originating from the two most popular communication channels associated with stable domestic political communication, #tcot (Top Conservatives on Twitter) and #p2 (Progressives 2.0). In total, this sampling procedure produced 1,522,415 tweets associated with Occupy Wall Street and 825,262 tweets associated with domestic political communication. As this analysis is concerned primarily with information spreading processes we consider only retweet events from this corpus, resulting in 676,369 retweets among 257,657 users associated with Occupy Wall Street, and 259,703 retweets among 68,049 users associated with stable domestic political communication. Henceforth, we consider these corpora to constitute representative samples of retweet interactions among users participating in the streams of content associated with the Occupy Wall Street movement and stable domestic political communication in the United States.

### Geocoding

To facilitate a geospatial analysis of communication activity associated with these content streams we require a high quality method to infer individual users' locations. To accomplish this, we rely on self-reported location strings and the services of a commercial geocoding API. This technique, popularized in work by Onnela et al. [34], has been shown to produce high-resolution, high-quality geolocation data in the presence of geographically meaningful input.

A caveat to this technique, however, is that it relies on raw text generated by a broad swath of the Twitter population, and so we find geographically meaningless location descriptors included in the dataset. To address this issue we rely on an extensive hand-curated blacklist of popular non-geographical responses such as 'everywhere' and 'the dance floor'. To produce this list we sorted all location strings by popularity and reviewed the thousand most popular strings manually, blacklisting those that did not correspond to geographically meaningful entities. Drawn from a long tailed distribution, 53% of all tweets in the data set are associated with a location among the 1,000 most popular responses, with 27% of all tweets containing one of the top hundred location strings. From this set of one thousand we blacklisted 161 non-location strings, corresponding to 6% of the tweets associated with the 1,000 most popular responses.

To improve recall in the presence of novel input, we used a modified version of the Ratcliff-Obershelp algorithm [35] to detect fuzzy matches between free text location strings and the blacklist of popular non-location responses. As a result, because 'the dance floor' is in the set of blacklist responses, strings taking a slightly modified form, such as 'on the dance floor,' will also be classified as invalid input. The hand-coded blacklist combined with the Ratcliff-Obershelp fuzzy matching technique resulted in 9% of the free-text location strings being classified as non-location input.

From among the remaining responses we submitted location strings to the Bing.com geocoding API, which returns a best-guess estimate for the corresponding physical coordinates. This output is hierarchically formatted to describe the finest level of geographic resolution available. For example, if a user reports 'Logan Square, Chicago' as his or her location, the Bing API will return information about the likely zip code, city, state and country associated with that location. However, if the user reports only 'USA,' the information provided by the API describes only a country-level guess as to the user's location. Owing to decreased coverage at the city-level and the proportionately few users associated with each individual city, we utilize the state-level location estimates for the geospatial components of this analysis.

In total, 68.4% of Occupy Wall Street users reported location strings, and from these we were able to obtain geolocation estimates for 55.7% of those accounts. Among this set of users, 60% of the resulting geolocation estimates included state-level metadata. Response rates were somewhat diminished for users associated with the stream of domestic political communication, with 36% of individuals reporting free-text location strings. Using the procedure described above, we were able to obtain geolocation estimates for 29.3% of all users in the domestic political communication stream, 82.4% of which contained state-level metadata.

## Geographic Profile

One of our goals is to establish a coarse-grained geographic profile for communication activity associated with the Occupy Wall Street movement. Formally, for each stream we define an activity distribution across states as, $A_i = \frac{T_i}{|T|}$, where $T_i$ is the total number of retweets originating from state $i$ and $|T|$ is the total number of retweets originating from all states. As outlined above, we focus on retweets as they correspond to attention allocation rather than total content production volume.

In addition to the distribution of activity across individual states we examine the information sharing relationships among users in different locations. To accomplish this, we rely on a network representation to characterize the flow of information on Twitter. Taking users as nodes, we define a weighted directed network in which an edge with weight $w$ is drawn from node $U_1$ to $U_2$ in the event that user $U_2$ retweets user $U_1$ $w$ times. The intuition underlying this approach is that each retweet provides evidence suggesting that information produced by user $A$ was evaluated and acted upon by user $B$.

Combining the user-level geocode metadata previously described with the network representation defined here we can induce another network describing the volume of communication between users in each state. In this network, nodes represent states, and weighted directed edges are drawn among them. The weight of the edge from $S_1$ to $S_2$ is defined as the sum of the weights among all edges originating from users in state $S_1$ and terminating in state $S_2$. We note, however, that this induced network must have geolocation labels for each node in a dyad. In the Occupy Wall Street stream we identify 143,437 tweets for which both the source and target have state-level geolocation data and 78,467 likewise restricted tweets in the stream of stable domestic political communication.

## Textual Content

Finally, we wish to investigate whether the content of tweets with different geospatial properties serve distinct communication functions. To accomplish this, we segregate Occupy Wall Street tweets into two classes: *interstate* tweets connect pairs of users in different states, and *intrastate* tweets connect users in the same state. We compute the probability of observing a token, $t$, in a tweet from a given class, $x$, as $P(t|x)$. Comparing these probabilities yields a ratio, $\frac{P(t|intrastate)}{P(t|interstate)}$, a value which is large when a token is more common in intrastate traffic than interstate traffic and small under the opposite conditions.

# Results

## Geographic Concentration

Figure 1, in which states are ordered according to the proportion of stream activity, shows that content in the Occupy stream is substantially more geographically concentrated in a few key states compared to domestic political communication. For example, New York accounts for 30% of the total retweet activity in the Occupy stream, while the most popular source for stable domestic political communication, Washington D.C., accounts for only 10.7% of the stream's total volume. As these plots make clear, the primary locations for on-the-ground Occupy activity are those places responsible for the majority of widely rebroadcast Occupy content, with California, New York and Washington D.C. acting as the source of 53.8% of total retweets. Figure 2 maps the states where the proportion of activity associated with the Occupy stream deviates the most from that associated with the stream of domestic political communication.

We also study the ratio of content production to content consumption by locale. Figure 3 shows this ratio, defined as the total number of retweets originating from users in that state divided by the total number of tweets retweeted by users in that state. This value serves to highlight the extent to which users in a given location are functioning as content producers or content consumers. Inspecting this plot, we find that in the Occupy stream users from just five states produced more content than they consumed. This stands in contrast to the stream of stable domestic political communication, in which fourteen states exhibit a ratio greater than one.

To highlight the effect of this geospatial concentration on communication flows between states it is instructive to visualize the structure of these networks. However, owing to the geographic aggregation process outlined in section *Geographic Profile* both networks are highly dense, with edges spanning most pairs of states. To address this issue we utilize a technique known as multiscale backbone extraction [36], which is useful for identifying statistically significant edges in weighted networks, regardless of the absolute value associated with the weight of that edge. This technique selects for edges with weights significantly above the expectation given by an analytically defined probability distribution that models a random allocation of each node's strength among its adjacent edges. Parameterized by a confidence level factor, $\alpha$, this technique allows for the selection of statistically significant edges across all weight scales, a feature that is especially valuable when working with networks with heterogeneous weight distributions such as those associated with communication or human mobility.

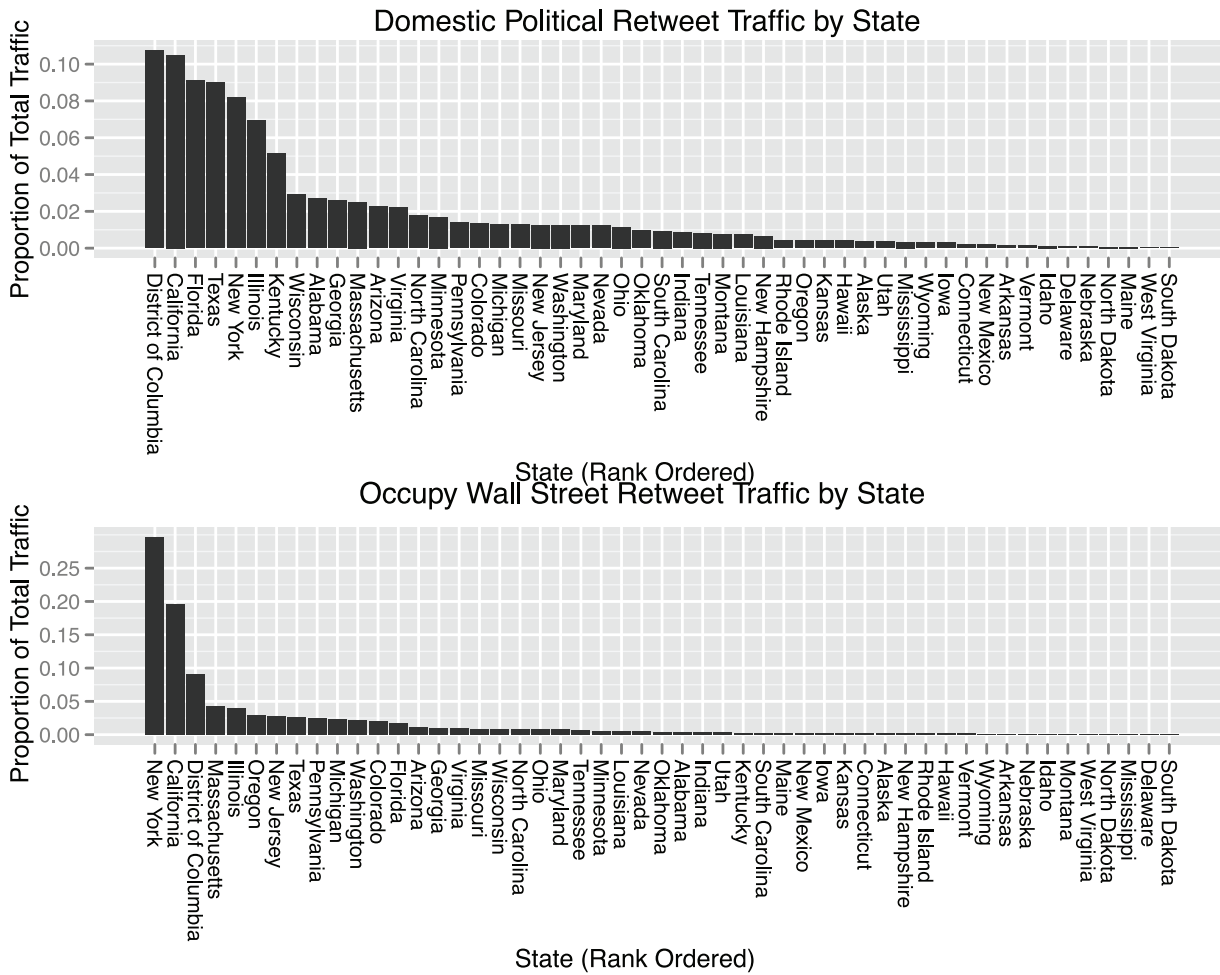

**Figure 1. Proportion of retweet traffic associated with each state, for each content stream.** Ordered by the amount of traffic associated with each state, it is clear that a few high-profile locations serve as the dominant sources of content in the Occupy stream. This concentration stands in contrast to the more heterogeneous activity profile for the stream of domestic political communication.
doi:10.1371/journal.pone.0055957.g001

Applying this technique to both networks reveals a communication backbone for the Occupy network that exhibits the highly concentrated hub and spoke structure described above. Figure 4 shows that the Occupy Wall Street network is characterized by minimal state-to-state connectivity, with the majority of statistically significant traffic flowing to and from New York, California and Washington D.C. This is in contrast to the communication backbone for the network of domestic political communication, in which we observe extensive interactions among many pairs of states.

## Localization

In Figure 5 we present interstate connectivity for each communication network as a matrix in which the weight of an edge is mapped to a grayscale hue ranging from white for weak relationships to black for the strongest relationships. Inspecting these plots, one of the most striking ways in which the topology of the Occupy Wall Street communication network departs from that of the domestic political communication network is the high degree of localization. This is evidenced by the presence of a strong diagonal in the Occupy Wall Street connectivity matrix, as well as the significant off-diagonal mass in the domestic political communication matrix. We find that 40% of Occupy retweets originate and terminate with users in the same state. In contrast, 11% of retweets from the domestic political stream exhibit this type of locality, an increase of more than 350%.

## Textual Analysis

To study the relationship between geography, resource mobilization, and collective framing, we focus on the content of tweets flowing within and between states. Restricting our analysis to tokens that account for at least 0.1% of both the intrastate and interstate tweet text, Table 1 presents the ten tokens most overrepresented in both intrastate communication as well as interstate communication.

## Discussion

The analysis of interstate connectivity patterns reveals that, relative to stable domestic political communication, the Occupy network has a highly localized geospatial structure, with a disproportionately large amount of traffic being produced and consumed by users in the same state. We propose that this phenomenon may be related to the issue of resource mobilization, that is, the process whereby any social movement must marshal resources such as money, infrastructure and human capital to further the goals of the movement. In the case of Occupy Wall Street, such resources are often quite tangible, and include not only tents and food, but also the participants required to facilitate large-scale protest action and extended encampments in cities across the country. In this light, it is easy to understand why such a disproportionately large fraction of attention is allocated to communication at the local level.

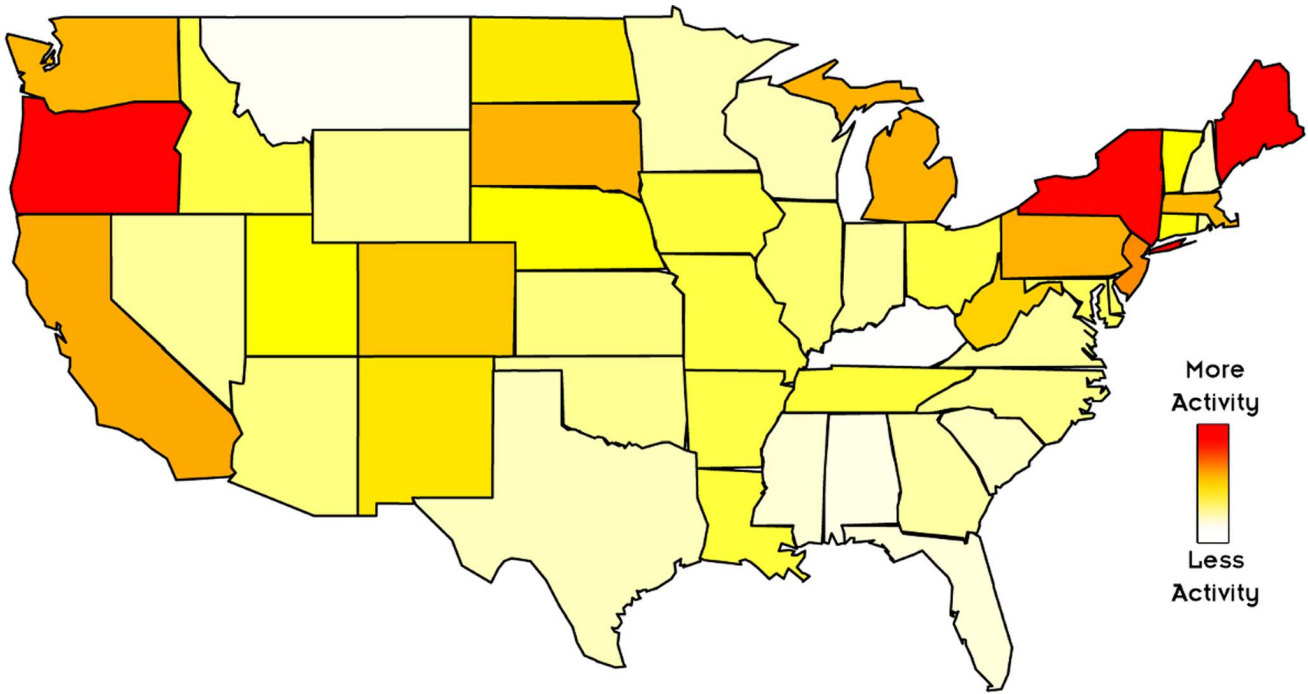

**Figure 2. Divergences in geographic distribution of users.** This cartogram uses color to represent the extent to which the number of Occupy Wall Street tweets in each state deviates from the domestic political communication baseline, computed as: $\frac{Occupy - Domestic}{Domestic}$. Redder colors indicate that proportionally more Occupy content originated from the associated state, while whiter colors indicate the opposite. To minimize the effect of outliers on the visualization and to highlight variation between states, colors for Maine and Oregon have been fixed, indicating that the deviation from baseline is more than three times the expected rate.
doi:10.1371/journal.pone.0055957.g002

With respect to the finding that the majority of widely rebroadcast content is produced by users in a small number of high profile locations, we observe that these states represent sites of major encampment and decision making activity. Despite the fact that all users can contribute equally to the Occupy stream, it appears that proximity to events on the ground plays a major role in determining which content receives the most attention. This is in contrast to the stream of domestic political communication, in which content from users across the United States is allocated a significant share of attention. Where the stream of domestic political communication looks more like a conversation taking place at the national level, the structure of the Occupy stream is more akin to a broadcast, with just a few locations playing the role of net content producers.

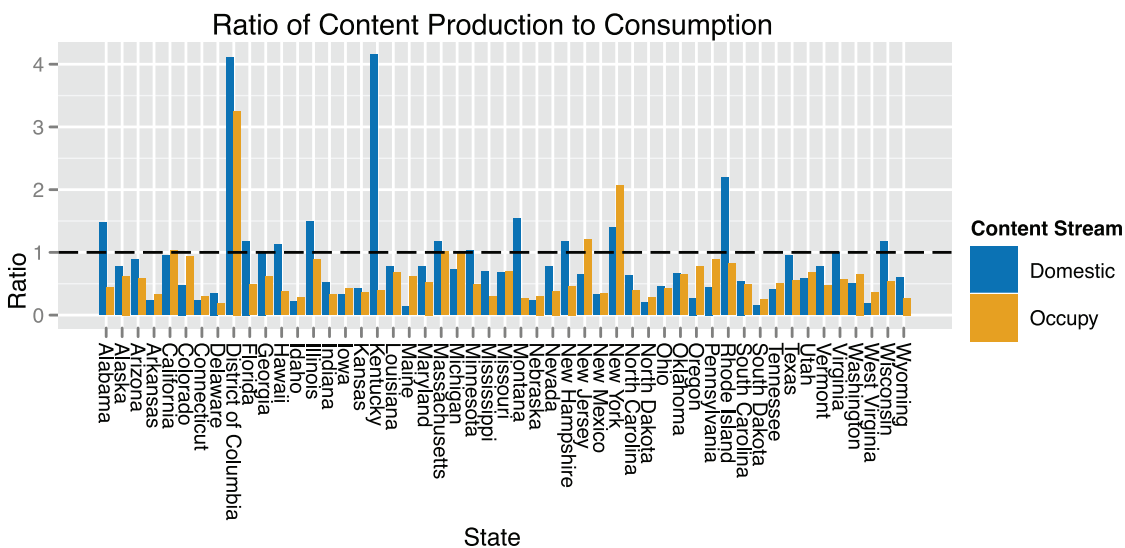

**Figure 3. Ratio of content production versus content consumption, by stream.** Occupy Wall Street users, by state, exhibit a lower content production to consumption ratio relative to users in the domestic political communication stream. The disproportionately high ratio observed for Kentucky can be attributed to the activity of a prolific, highly popular left-leaning user from that state.
doi:10.1371/journal.pone.0055957.g003

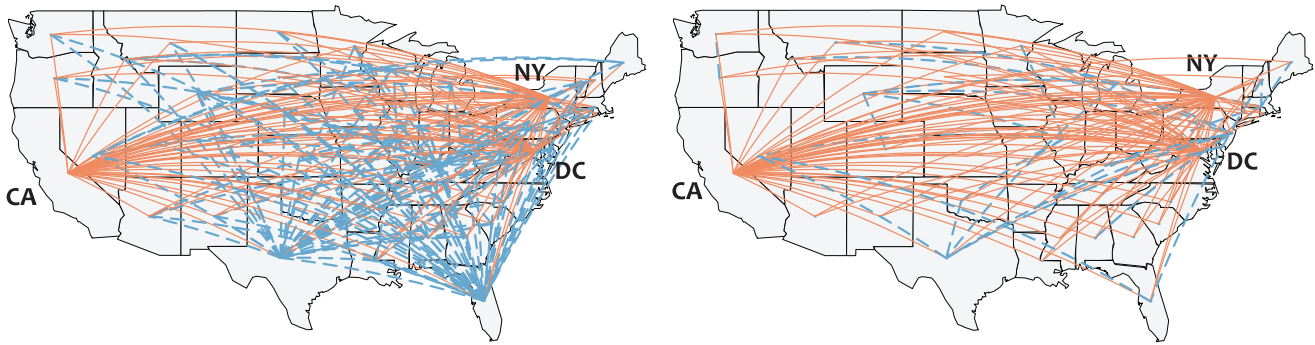

**Figure 4. Multiscale backbone (α = .15) of the continental interstate communication networks.** Stable domestic political communication is shown at left, Occupy Wall Street at right. Edges adjacent to New York, California, and Washington D.C. are shown in red. Note that Occupy Wall Street exhibits a clear hub-and-spoke pattern, with the majority of traffic flowing to and from New York, California, and Washington, D.C. Likewise, observe that the Occupy Wall Street network exhibits diminished levels of interstate connectivity compared to the network of domestic political communication. We note that the structure of this network backbone is robust to different parameterizations of alpha.
doi:10.1371/journal.pone.0055957.g004

Finally, we propose that interstate communication plays a significant role in the propagation of narrative imagery associated with collective framing processes, and that intrastate communication is driven more predominantly by the pressures of resource mobilization. Looking to the lists of tokens most overrepresented in each type of traffic (Table 1), we find that those more common in interstate communication include references to core framing language and the news media. This finding suggests that when users engage in communication across state boundaries they allocate proportionately higher levels of attention to speech associated with collective framing processes. In contrast, the tokens more common in intrastate traffic relate to protest action and specific times and places. From this we conclude that the

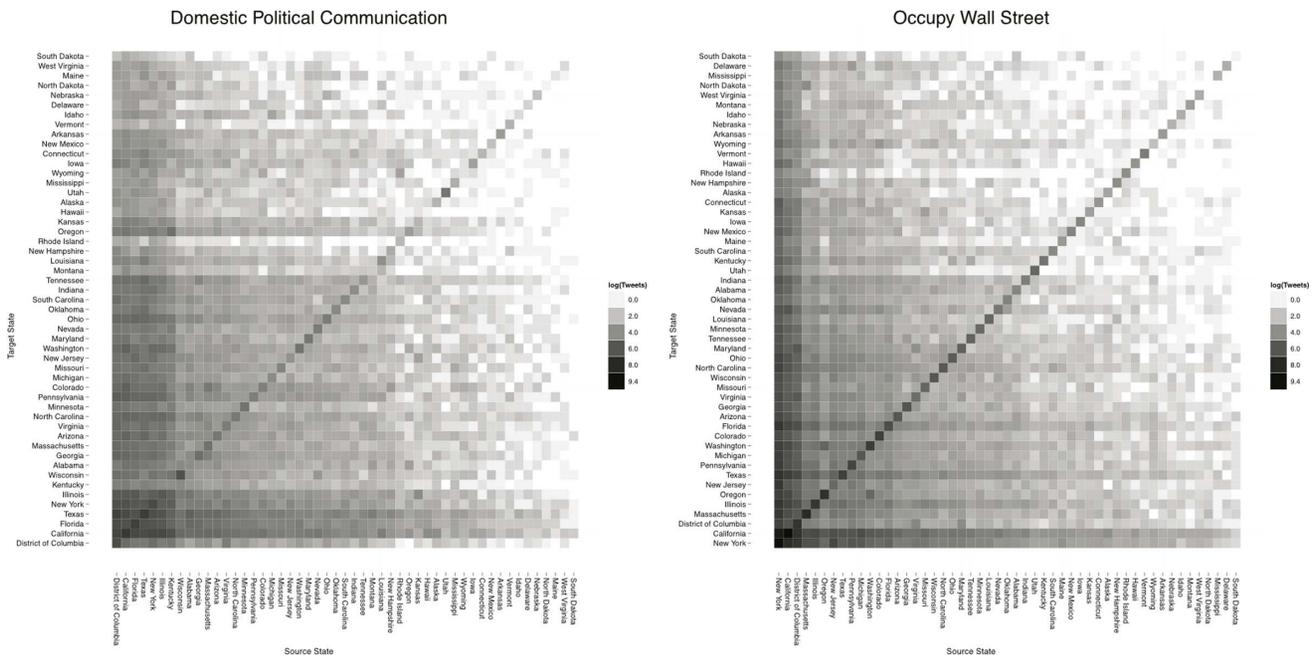

**Figure 5. Connectivity matrices describing directed interstate communication volume.** The edge weight corresponding to each cell is mapped to a color hue on a logarithmic scale ranging from white for edges with the least weight to black for edges with the most weight. The strong diaonalization and limited off-diagonal mass apparent in the Occupy Wall Street matrix is indicative of highly localized communication activity.
doi:10.1371/journal.pone.0055957.g005

**Table 1.** Lists of tokens most overrepresented in intrastate and interstate communication.

| Interstate | | Intrastate | |
|---|---|---|---|
| Token | Ratio | Token | Ratio |
| wall | .590 | city | 2.254 |
| nyc | .600 | tonight | 1.737 |
| street | .699 | march | 1.669 |
| news | .718 | join | 1.494 |
| 99% | .756 | solidarity | 1.387 |
| bank | .763 | day | 1.354 |
| don't | .782 | square | 1.333 |
| media | .837 | please | 1.243 |
| peaceful | .845 | park | 1.220 |
| nypd | .847 | now | 1.179 |

'Ratio', defined as $\frac{P(Token|Intrastate)}{P(Token|Interstate)}$ is small when a token is more common in intrastate traffic and large when a token is more common in interstate traffic. Terms relating to rallying supporters are more predominant in intrastate communication, while interstate traffic tends to favor terms such as protest slogans and references to the media.
doi:10.1371/journal.pone.0055957.t001

content of intrastate tweets deals much more frequently with rallying the movement's participants, a core function of resource mobilization.

The findings outlined in this paper dovetail nicely with established literature on social movement theory. However, statistical measures are limited in the extent to which they can accurately represent nuanced features of communication, and future work in this domain would benefit from rigorous qualitative content analysis. Moreover, there remains room to improve our understanding of how closely the structure of social media communication mirrors that of other forms of communication. For example, Mislove, et al. found that the geographical distribution of Twitter users tends to over-represent populous counties and metropolitan areas, suggesting that entire rural regions may be significantly under-represented – with similar findings holding true for ethnicity and gender as well [37]. In this respect as well, work of this nature would benefit from deeper involvement from scholars in the social sciences, and we hope that this type of interdisciplinary collaboration will become increasingly common.

## Acknowledgments

We would like to thank Alex Rudnick, Jacob Ratkiewicz, Mark Meiss, and other current and past members of the Truthy group at Indiana University (cnets.indiana.edu/groups/nan/truthy) for their contributions to the Truthy Project. Additionally, we would like to thank Fabio Rojas, Brian Keegan, and Bruno Gonçalves for their constructive, insightful feedback during the preparation of this manuscript.

## Author Contributions

Conceived and designed the experiments: MDC AF FM EF CD KM. Performed the experiments: MDC CD. Analyzed the data: MDC AF FM EF CD KM. Wrote the paper: MDC AF FM EF.